\documentclass[aps,twocolumn,showpacs,prl,floatfix]{revtex4}
\usepackage{graphicx}
\usepackage{epsf}
\usepackage{wrapfig}
\usepackage{epsfig}
\usepackage{sublabel}
\usepackage{epsfig}
\newcommand{\be}{\begin{eqnarray}}
\newcommand{\ee}{\end{eqnarray}}
\newcommand{\beq}{\begin{eqnarray}}
\newcommand{\eeq}{\end{eqnarray}}

\newcommand{\spettN}{P_N^A(k,E)}
\newcommand{\spett}{P^A(k,E)}

\newcommand{\sez}{\frac{d^{2}\sigma(q,\nu)}{d{\Omega}_2\,d{\nu}}}
\newcommand{\sezdue}{\sigma_2^A (q,\nu)}
\begin{document}
%\today
%\vskip 2mm \date{\today}\vskip 2mm
\title{ Obtaining information on short-range correlations from inclusive electron scattering}
\author{Claudio Ciofi degli Atti, Chiara Benedetta Mezzetti}
\address{
Department of Physics, University of  Perugia and  Istituto
Nazionale di Fisica Nucleare,\\ Sezione di Perugia, Via A.Pascoli,
I-06100 Perugia, Italy }
%\date{\today}
\begin{abstract}
 In view of recent data from the Thomas Jefferson National Accelerator Facility (JLab) on inclusive electron scattering off nuclei at high
 momentum transfer ($Q^2\gtrsim \, 1\,GeV^2$)
 and their current analysis, it  is shown that,  if the scaling
 variable is properly chosen,
 the analysis  in terms
 of scaling functions
  can provide useful
 information on short-range correlations (SRC).
This is demonstrated by introducing  a new relativistic scaling
variable that incorporates the momentum dependence of the
excitation energy of the $(A-1)$ system,  with the resulting
scaling function being closely related to  the longitudinal
momentum distributions.
\end{abstract}
 \maketitle
%\narrowtext
\newpage
Obtaining information on short-range correlations SRC in nuclei is
a primary goal of modern nuclear physics \cite{quarkcosmos}.
Interest in  SRC  stems not only from the necessity to firmly
establish the limits of validity of the standard model of nuclei
but also from the impact that the knowledge of the detailed
mechanism of SRC would have  in understanding  the role played by
quark degrees of freedom in hadronic matter  and   the properties
of the latter in dense configurations \cite{review}. Recently,
evidence of SRC has been provided by new experimental data on
inclusive [$A(e,e')X$] \cite{ratioAD,ratioA3} and exclusive
[$A(e,e'pN)X$ and $A(p,pN)X$] lepton and hadron scattering  off
nuclei at high momentum transfer ($Q^2 \gtrsim 1$ $GeV^2$) (see
Ref. \cite{science} and references therein quoted). In inclusive
scattering
 the observation  of a scaling behavior of
the ratio of the cross section on heavy nuclei to that on the
deuteron \cite{ratioAD}, for values of the Bjorken scaling
variable $1.4 \lesssim x_{Bj} \lesssim 2$, and to that on  $^3He$
\cite{ratioA3},  for $2 \lesssim x_B \lesssim 3$, has been
interpreted as evidence that the electron probes two- and
three-nucleon correlations in complex nuclei similar to the ones
occurring in two- and three-nucleon systems \cite{mark1}. It
should be pointed out, however,  that whereas exclusive processes
can directly access the relative and center-of-mass motions of a
correlated pair in a nucleus \cite{theoryexclusive}, obtaining
information on these quantities from inclusive scattering is, in
principle,  more difficult.  Various approaches based on scaling
concepts have  therefore been proposed, going from the already
mentioned scaling behavior of the cross section ratio plotted
versus  $x_{Bj}$, to the scaling behavior of the ratio of the
nuclear to the nucleon cross sections plotted versus proper
scaling variables; among the latter, a process that has been most
investigated in the past is the so-called Y-scaling, for it is
believed that this may  represent a powerful  tool to extract the
high-momentum part of the
 nucleon momentum distribution
which is governed by  SRC \cite{ciofi,day}. It is the aim of this
Rapid Communication to critically reanalyze the concept of Y-scaling,  mainly
because of i) the lack of a general consensus about the
usefulness of such a concept and ii) a strong renewal of interest
in Y-scaling owing to recent
  experimental data on $A(e,e')X$ reactions from the Thomas Jefferson National Accelerator Facility (JLab) \cite{arrington,newdata}.
  We will
 show that the
 analysis of inclusive scattering in terms of proper Y-scaling variables
 could indeed provide  useful   information on
 SRC; to this
 end, following the suggestion of Refs. \cite{CW,CFW1,CFW2} a
new approach to Y-scaling and its usefulness will be illustrated
in detail.  Let us consider a  virtual photon of high momentum
impinging on a nucleus $A$ (with mass $M_A$) and knocking out, in
a
 quasielastic
process,
  a nucleon $N$ (with mass $m_N$) having
momentum $k\equiv |{\bf k}|$ and removal energy $E$.
The latter is defined as the energy necessary to remove the nucleon
 from $A$
leaving the residual nucleus $(A-1)$ (with mass $M_{A-1}$) with
intrinsic excitation energy $E_{A-1}^*$ (i.e., $E=m_N+M_{A-1}- M_A
+E_{A-1}^*=E_{min}+E_{A-1}^*$).
 In Plane Wave Impulse Approximation (PWIA) and using  the instant form
 of dynamics,
  the quasielastic cross-section reads as follows
%%*******************************************************************EQ(X-section)
\begin{widetext}
\be
 \sigma_2^A (q,\nu)\equiv \sez
 = \sum_{N=1}^{A} \int d\,E\,d\, {\bf k} \spettN
 \sigma_{eN}(q,\nu,{\bf k},E)\,\delta(\nu+M_{A} -E_N-E_{A-1})
 \label{X-section}
\ee
 \end{widetext}
 %%*******************************************************************************
%%%
\noindent with energy conservation  $(M_{A-1}^*=M_{A-1}+E_{A-1}^*)$
%%************************************************************************EQ(energycons)
 \be
\nu+M_{A} =\sqrt{m_N^2+({\bf k}+{\bf q})^2}
+\sqrt{{M_{A-1}^{*^{2}}} +{\bf k}^{2}}
\label{energycons}
\ee
%%%********************************************************************************
\noindent and momentum conservation  ${\bf q}={\bf p}+{\bf
p}_{A-1}$. Here $\nu = \epsilon_1 -\epsilon_2$ and ${\bf q}= {\bf
k}_1 -{\bf k}_2$ are the energy
 and three-momentum transfers
 ($Q^2=q^2-\nu^2=4\epsilon_1\,\epsilon_2 \sin^2 \frac{\theta}
{2}$ with
 $q\equiv |{\bf q}|$),
$\sigma_{eN}$ is the elastic
electron cross section off a moving off-shell nucleon with
 momentum $k\equiv |{\bf k}|$ and removal
energy $E$, and $P_N^A(k,E)$ is the  spectral function (normalized
to one) of nucleon $N$ (i.e., the joint probability to have a
nucleon with momentum $k$ and removal energy $E$); eventually,
${\bf p}$ and ${\bf P}_{A-1}$ are the momenta of the
 undetected
struck nucleon and the final $(A-1)$ system. Considering,
 for ease of presentation,
 isoscalar nuclei,
one has $P_N^A(k,E)=P_p^A(k,E)=P_n^A(k,E)\equiv
  P^A(k,E)=P_0^A(k,E)+P_1^A(k,E)$,
  where $P_0^A(k,E)=(1/A)\sum\displaylimits_{\alpha {\in F}}
  A_\alpha n_\alpha(k)\,\delta(E-\epsilon_\alpha)$ is the
  (trivial) shell-model part
  [with $A_\alpha$ denoting the occupation number
  of the single-particle state $\alpha$ with removal
   energy $\epsilon_\alpha$ and momentum
  distribution $n_\alpha(k)$],
and $P_1$ is the (interesting) part generated by NN correlations.
The spectral function is linked to the momentum distributions by the momentum sum
rule $n^A(k)=\int P^A(k,E)\,dE=
\int P_0^A(k,E)\,dE+\int P_1^A(k,E)\,dE=
 n_0^A(k)+n_1^A(k)$.
It has  been shown
 \cite{ciofi} that  at high values of
momentum transfer, after integrating over the direction of ${\bf
k}$,  Eq. (\ref{X-section})  can be written, to a good
approximation,  as follows:
%%%******************************************************************EQ(crossappr)
\begin{widetext}
 \be
 \sezdue\simeq \left\{
 [Zs_{ep}(q,\nu,k, E)+Ns_{en}(q,\nu,k, E)]
\frac{E_{p}}{q} \right\}_{(k_{min},E_{min})}  \times F^A(q,\nu)
\label{crossappr}
 \ee
 \end{widetext}
%%%*****************************************************************************
where $s_{eN}$ is the electron-nucleon cross section integrated
over the polar angle, and  $F^A(q,\nu)$ is the nuclear structure
function
%%%*************************************************************EQ(scalingfun)
\be
F^A(q,\nu)=2\,\pi\,
\int\displaylimits_{E_{min}}^{E_{max}(q,\nu)}
d\,E\int\displaylimits_{k_{min}(q,\nu,E)}^{k_{max}(q,\nu, E)}
k\,d\,k\, \spett
\label{scalingfun}
\eeq
%%%*******************************************************************************
Eq. (\ref{crossappr})  is obtained by eliminating the
$\delta$ function  by integrating
 over $\cos\alpha=({{\bf k}\cdot {\bf q}}/kq)$,
 with  the limits of integration resulting from the condition
$-1 \leq \cos\alpha \leq 1$.
  We can now introduce a scaling variable $Y=Y(q,\nu)$, which is only required
 to be a function of
$q$ and $\nu$ (and any arbitrary constant)  so that, no matter
with the specific form of $Y$, the cross section and  the
structure function can be expressed
 not in terms
of the two canonical independent variables $q$ and $\nu$, but,
 without loss of generality,  in terms of
$q$ and $Y=Y(q,\nu)$. Correspondingly, a  scaling function
 $F^A(q,Y)$ is introduced; this is nothing but Eq. (\ref{scalingfun}) with  $\nu$ replaced
 everywhere by $Y$; if, under certain conditions,
$F^A(q,Y)\rightarrow F^A(Y)$, Y-scaling is said to occur and,
depending on the physical meaning of $Y$ and $F^A(Y)$, various
information on nucleons in nuclei could be obtained. To simplify
our analysis, let us consider
 high values of the momentum transfer,  when  $E_{max}(q,Y)$ and
  $k_{max}(q,Y,E)$
 become so large that, because of the rapid falloff fall-off of
 $P^A(k,E)$,  they can be replaced by $\infty$
(although in actual calculations we use the correct values of these
quantities); in this case,  the $q$ and $\nu$ dependence of the
scaling
 function is governed  only by $k_{min}(q,Y,E)$, and it is  trivial to
 show  that, by adding and subtracting a proper term,
  the scaling function can be cast in the following general form
 %%%%**************************************************************************EQ(fqy)
\begin{widetext}
\be
F^A(q,Y)= 2\pi\int\displaylimits_{E_{min}}^{\infty}
d\,E\int\displaylimits_{k_{min}(q,Y,E)}^{\infty}
k\,d\,k\, \spett = f^A(Y) -B^A(q,Y)
\label{fqY}
\ee
\end{widetext}
%%%%***************************************************************************************
\noindent where
$f^A(Y)=2\pi\int\displaylimits_{|Y|}^{\infty}k\,d\,k\, n^A(k)$
%%%******************************************************************************EQ(ennelong)
% \begin{widetext}
%\be f^A(Y)=
%2\pi\int\displaylimits_{|Y|}^
%{\infty}kdk\int\displaylimits_{E_{min}}^{\infty} d\,E \,\spett=
%2\pi\int\displaylimits_{|Y|}^{\infty}k\,d\,k
%\,n^A(k)
%=2\pi\int\displaylimits_{0}^ {\infty}d\,{\bf
%k_{\perp}}n^A({\bf k_\perp},k_{||}=Y) \label{ennelong}
%\ee
%\end{widetext}
%%%******************************************************************************************
%\noindent
represents the {\it longitudinal momentum distribution},
and
%%%****************************************************************************EQ(by)
\be
B^A(q,Y) = 2 \pi \int\displaylimits_{E_{min}}^{\infty}d\! E
            \int\displaylimits _{\vert Y\vert}^{k_{min}(q,Y,E)}\, k\,d\! k \, P_{1}^A (k,E)
            \label{by}
\ee
%%****************************************************************************************
is the {\it binding correction} \cite{ciofi}, which,
through $ k_{min}(q,Y,E)$, is governed by the continuum energy
spectrum of the final $(A-1)$ system, unlike  $f^A(Y)$, which is
integrated over all excited states of $(A-1)$. The quantities
  $f^A(Y)$ and  $n^A(k)$
 are linked by the relation
$n^A(k)=[df^A(Y)/dY]/[2\pi\,Y],  k=|Y|$,  so that if $f^A(Y)$
could be extracted from the experimental data, $n^A(k)$ could be
determined. Unfortunately, such an extraction is hindered by  the presence of $B^A \neq 0$ depending upon
 the
difference between $Y$ and $k_{min}$ and therefore
upon the definition of the
former.
The binding correction
 is absent only in the deuteron, since $E=E_{min}=cost=2.22 \,MeV$, so that
 $Y=k_{min}(q,\nu,E_{min})$, $B^D(q,Y)=0$ and
 $F^D(q,Y) = f^D(Y)$. The final state interaction  (FSI) of the struck nucleon
 invalidates the PWIA, but in spite of that,  an approach was developed in the past to reduce
the effects from both the binding corrections and FSI
\cite{ciofi};
 the approach
is based upon the  widely used relativistic scaling variable
  $Y=y$ \cite{ciofi,day,arrington,newdata,poly}, which is obtained  by setting
  in the energy
conservation equation [Eq. (\ref{energycons})] $k=y$, ${{\bf k}\cdot {\bf
q}}/{kq} = 1$  and, most importantly, $E_{A-1}^{*}=0$; thus  $y$
represents the {\it minimum longitudinal momentum of a nucleon
having the minimum value of the removal energy  $E=E_{min}$}. In
the asymptotic limit ($q \to \infty$),
 Eq. (\ref{fqY}) scales in $y$
and becomes the {\it asymptotic scaling function}   $F^A(y)=f^A(y)
-B^A(y)$, that is Eq. (\ref{fqY}) with $Y$ and $k_{min}(q,Y,E)$
replaced by $y$ and $k_{min}^\infty(y,E)$, respectively
%%%***************************************************************************EQ(asfy)
%\begin{widetext}
%\beq
%F^A(y)=2\pi\int\displaylimits_{E_{min}}^{\infty}
%d\!E\int\displaylimits_{k_{min}^\infty(y,E)}^{\infty}
%k\,d\,k\, \spett
%=f^A(y) -B^A(y)
%\label{asfy}
%\eeq
%\end{widetext}
%%%************************************************************************************
 (scaling in this variable
 also occurring within a relativistic description of the deuteron
 \cite{poly}).
 Unfortunately, owing to the presence of
  $B^A(y)$, $F^A(y)$ is not
 related to a momentum distribution so that,  in principle, the experimental longitudinal
 momentum distribution
$f_{ex}^A(y)$ and, consequently, $n_{ex}^A(k)$, cannot
be extracted from the data. Let us briefly
 recall how this problem was addressed in Ref. \cite{ciofi}.
 The experimental scaling function
%%%********************************************************************************EQ(effeexp)
%\begin{widetext}
%\be
$ F_{ex}^A(q,Y) =
%\frac{d^2{\sigma}^{ex}(q,Y)}{d\Omega_2\,d\nu}}
\sigma_{2,ex}^{A}(q,Y) /\{[Zs_{ep}(q,\nu,k, E)+Ns_{en}(q,\nu,k,
E)] (E_{p}/q)\}_{(k_{min},E_{min})}$
%\label{effeexp}
 %\ee
 %%%***************************************************************************************
 %\end{widetext}
%\noindent
 exhibits, when $Y=y$, a strong $q$ dependence owing to the FSI and binding effects and
differs from the asymptotic scaling function
$F_{ex}^A(y)$.  The latter, however, has been
 obtained in Ref. \cite{ciofi}
 by
extrapolating to $q \rightarrow \infty$ the available values
of $F^A_{ex}(q,y)$, on the basis that FSI can be represented as a
 power series in $1/q$  and dies  out at large $q^2$,
 a conclusion that has been reached by
  various authors (see e.g. Refs. \cite{rinawest}).
 The experimental
longitudinal  momentum distribution $f_{ex}^A(y)$ has thereby been
obtained by adding to $F_{ex}^A(y)$ the binding correction
$B^A(y)$ evaluated theoretically, and $n_{ex}^A(k)$ has been
obtained by $n^A(k)=-d[F^A(y)+B^A(y)/dy]/[2\pi\,y]$, $k=|y|$. Such
a
 procedure affects the final results in terms of large errors on
 the extracted
momentum distributions, particularly at large values of $k$; in spite of
 these errors,
the extracted momentum distributions at $k\gtrsim
1.5-2\,\,fm^{-1}$ turned out to be larger  by orders of magnitude
from the prediction of mean-field approaches, and in qualitative
agreement with realistic many-body calculations that include SRC.
To make  the
   extraction of $f_{ex}^A(y)$ as independent as possible
   from theoretical binding
   corrections, in Ref.
   \cite{CW} another scaling variable $Y=y_{CW}$ has been introduced; this scaling variable
    incorporates relevant physical dynamical effects left out
   in the definition of $y$. To readily understand the physical meaning
   of the new scaling variable,  let
   us consider the asymptotic limit of $k_{min}(y,q, E)$
   for a large nucleus [i.e.,
   $k_{min}^{\infty}(y, E) = |y-(E-E_{min})|$]; it can be seen  that
    only when $E=E_{min}$ does
$k_{min}(y, E) = \vert y \vert$, in which case  $B^A= 0$ and
$F^A(y)=f^A(y)$;  this holds only for the deuteron, whereas for a
complex nucleus  $E_{A-1}^{*} \ne 0$ and $E \geq E_{min}$, so
$B^A(y) \ne 0$, and $F^A(y) \ne f^A(y)$. It is therefore the
dependence of $k_{min}$ on $E_{A-1}^{*}$ that gives rise to the
binding effect [i.e., to the relation  $F^A(y) \ne f^A(y)$]. This is
an unavoidable defect of the usual approach to Y-scaling, based on
the scaling variable $y$; in fact, the longitudinal
momentum is very different for weakly bound, shell-model nucleons
(${E_{A-1}^*} \sim 0-20\, MeV$) and  strongly bound, correlated
nucleons (${E_{A-1}^*} \sim 50-200 \,MeV$), and  at large values
of
 $|y|$ the scaling
function is not related to the longitudinal momentum of strongly
bound  correlated nucleons,  whose contributions almost entirely
exhaust the behavior of the scaling function. As stressed in Refs.
\cite{CW,CFW1,CFW2}, to establish a global link between
experimental data  and longitudinal momentum components, one has
to conceive a scaling variable {\it that could equally well
represent longitudinal momenta of both weakly bound and strongly
bound nucleons so that the binding correction could be minimized}.
 %%%%%%%%%%%%%%%%%%%%%%%%%%%%%%%%%%%%%%%%%%%%%%%%%%%%%%%%%%%%%%%%%%%%%%%%%%%%%%%%%%%%%%% FIG.1
\begin{figure}[!hbp]
%\vskip 0.5cm
\centerline{\centerline{\epsfig{file=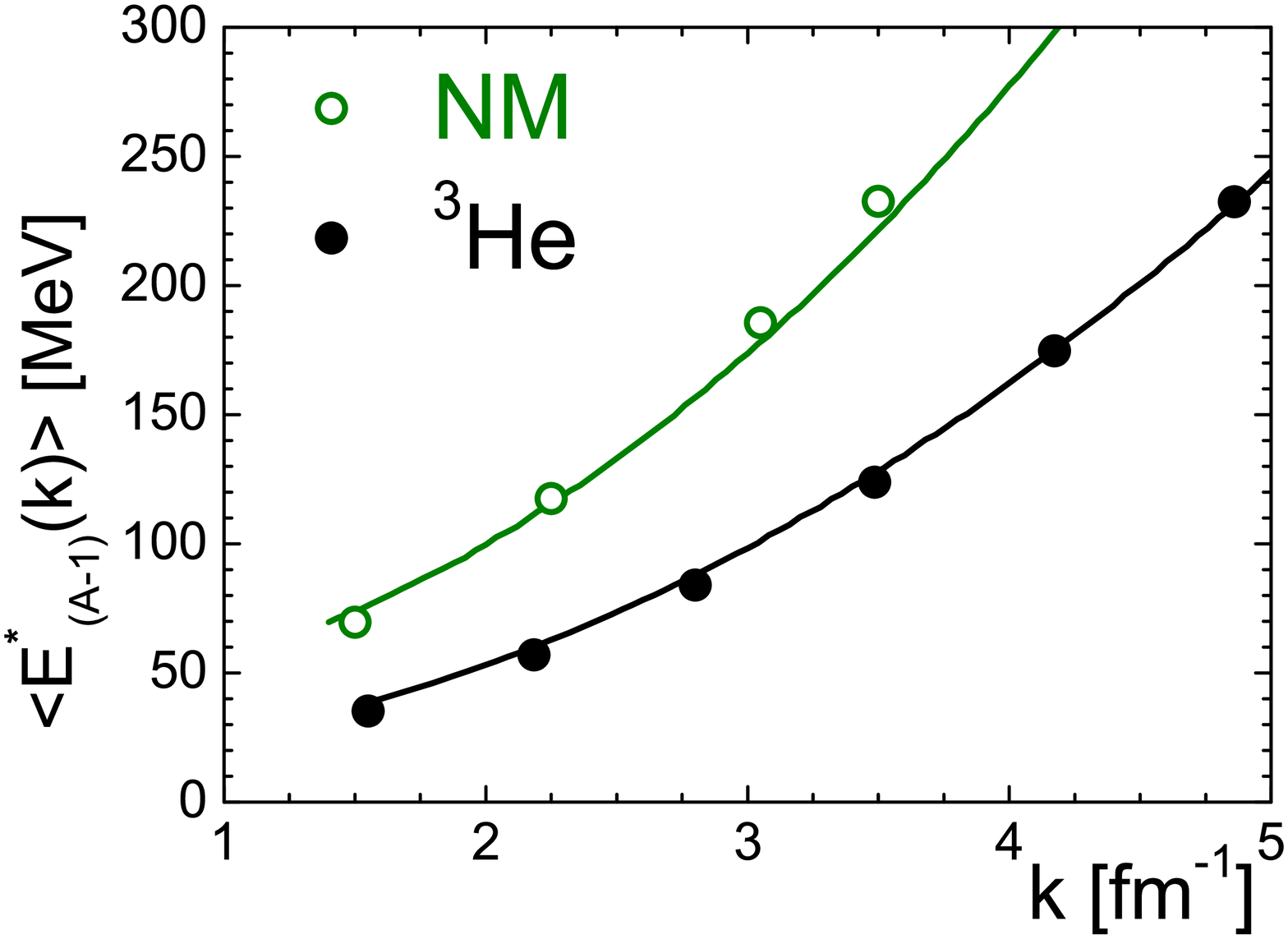,width=9.7cm,height=5.5cm}}}
\vskip -0.3cm \caption{The average value of $E_{A-1}^*(k)$ [Eq.
(\ref{average})] calculated for nuclear matter with the spectral
function of Ref. \cite{panda} (open dots), and for $^3He$ with the
spectral function from the Pisa wave functions \cite{pisa} (full
dots). The full lines are obtained with the spectral function of
the few-nucleon correlation model of Ref.
\cite{ciosim}.}\label{Fig1}
\end{figure}
%%%%%%%%%%%%%%%%%%%%%%%%%%%%%%%%%%%%%%%%%%%%%%%%%%%%%%%%%%%%%%
%%%%%%%%%%%%%%%%%%%%%%%%%%%%%%%%%%%%%%%%%%%%%%%%%%%%%%%%%%%%%%
This can be achieved  by adopting a
 scaling variable that
properly includes  the momentum dependence of the average
excitation energy  of $(A-1)$ generated by correlations, namely,
%%%%**********************************************************************EQ{average}
\beq
<E_{A-1}^*(k)>= \frac{1}{n^A(k)}\int P_1^A(k,E_{A-1}^*)
E_{A-1}^*d\,E_{A-1}^*
\label{average}
 \eeq
%%%*********************************************************************************
\noindent where $E_{A-1}^*=E-E_{thr}^{(2)}$,\,
$E_{thr}^{(2)}=M_{A-2}+2m_N-M_A$ being the threshold energy for
two-particle emission.
 We have calculated the quantity in Eq. (\ref{average})
  using  a realistic  spectral function for nuclear matter and
 $^3He$.  The results are
 presented in Fig. \ref{Fig1}, where they are  compared
with the prediction of  the spectral function of
  the few-nucleon correlation (FNC) model
   of Ref. \cite{ciosim}, according to which
 %%*****************************************************************************EQ(twelve)
 \begin{equation}
E_{A-1}^{*}({\bf k}, {\bf K_{CM}})= \frac{A-2}{A-1}{\frac{1}{2m_N}}\left[{\bf {k}} -\frac
{A-1}{A-2}{\bf {K}}_{CM}\right]^2
\label{twelve}
 \end{equation}
 %%**************************************************************************************
 \noindent where ${\bf {K}}_{CM}$ is the CM momentum of a correlated pair.
In view of the very good
 agreement between the FNC model and the exact many-body results for
 nuclear matter and $^3He$, we used
 the former
 to calculate $<E_{A-1}^*(k)>$ for nuclei with $3< A <\infty$.
 The values shown  in Fig. \ref{Fig1}
 can be interpolated  by
 %%***********************************************************************EQ(tredici)
 \begin{equation}
<E_{A-1}^{*}(k)> = \frac{A-2}{A-1}T_N
+b_A-{c_A}|{\bf k}|
\label{tredici}
 \end{equation}
 %%************************************************************************************
  \noindent where  $T_N=(\sqrt{m_N^2+k^2}-m_N)$, and $b_A$ and $c_A$
 result from the CM motion of the
pair ($b_{NM}=37.3\,MeV$, $c_{NM}=0.04$ and $b_{3}=-2.94\,MeV$,
$c_{3}=-0.03$). Placing in  Eq. (\ref{energycons})
 $k=y_{CW}$,  $\frac{{\bf
k}\cdot {\bf q}}{kq} = 1$ and  $M_{A-1}^*= M_{A-1} +
<E_{A-1}^{*}(k)>- <E_{gr}>$,   we obtain a  fourth-order equation
 for the new scaling
variable $y_{CW}$, which,  in contrast to
previous work \cite{CW,CFW1,CFW2}, has been solved  exactly;
%%%%%%%%%%%%%%%%%%%%%%%%%%%%%%%%%%%%%%%%%%%%%%%%%%%%%%%%%%%%%%%%%%%%%%%%%%%%%%%%%%%%%%% FIG.2
\begin{figure}[!htp]
  \vskip 0.2cm \centerline{
   \hskip -0.5cm
\epsfig{file=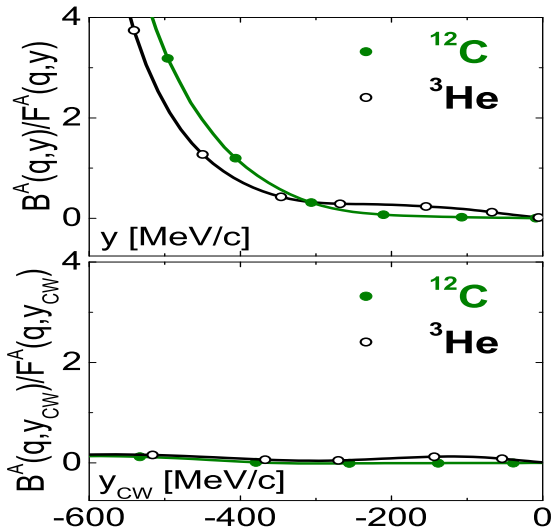,width=12cm,height=7.0cm}}
 \vskip 0.3cm \caption{The ratio of the  binding correction
  [Eq. (\ref{by}] to the scaling function [Eq. (\ref{fqY})] for $^3He$
  (open dots)
  and $^{12}C$ (full dots) calculated with the scaling variable $y$,
   which does
  not contain any effective excitation energy from SRC (upper panel),
  and with the variable
  $y_{CW}$, which takes
  into account SRC effects by Eq. (\ref{tredici}) (lower panel).}
  \label{Fig2}
\end{figure}
this, together with the relativistic extension of the definition
of the mean excitation energy, is necessary to extend $y_{CW}$ to
high values. Note that the value of  $<~E_{gr}>$, fixed by  the
Koltun sum rule (see Refs. \cite{CW,CFW1,CFW2}),  has been added
to Eq. (\ref{tredici})
 to counterbalance the effects of $<E_{A-1}^*>$ at low  $y_{CW}$. For a
large nucleus and not too large values of $y_{CW}$, one has
%%*************************************************************** EQ(quattordici)
\be
y_{CW}= -\frac{\tilde q}{2} + \frac{\nu_A}{2W_A}
\sqrt{{W_A^2}-{ 4m_N^2}}
\label{quattordici}
 \ee
%%*******************************************************************************
%
Here, $\nu_A = \nu +  \tilde M_D$, $\tilde M_D = 2m_N -
E_{th}^{(2)} - b_A + <~E_{gr}>$, $\tilde q = q+c_A{\nu}_A$ and
$\rm{W}_{A}^{2} = {\nu_A}^2 - {\bf q}^2 =\tilde M_D^{2} + 2 \nu
\tilde M_D - Q^{2}$. For the deuteron $y_{CW}= y = -{q}/{2} +
(\nu_D/2W_D) \sqrt {W_D^2 - {4 m_N^2}}$
 with ${\nu_D} =
\nu + M_{D}$ and  invariant mass ${\rm W}_{D}^{2} =  {\nu_D}^2 -
{\bf q}^2 = {M_D}^2 + 2{\nu}M_D- Q^2$;   for small values of
$y_{CW}$, such that ${\frac{A-2}{A-1}(\sqrt{y_{CW}^2+m_N^2}-m_N)
+b_A-{c_A}{|y_{CW}|}} \ll <E_{gr}>$, the  variable $y$,
representing the longitudinal momentum of a weakly bound nucleon,
is recovered. Therefore $y_{CW}$ effectively takes into account
the $k$ dependence of  $E_{A-1}^*$, both at low and high values of
$y_{CW}$, and  interpolates between the correlation and the single-particle regions;
it   can be interpreted as the {\it minimum
longitudinal
 momentum of a nucleon that, at high values of $y_{CW}$,  has removal energy $<E_{A-1}^*>$
and is partner of a  correlated two-nucleon pair with effective mass
$\tilde M_D$.}

%%%%%%%%%%%%%%%%%%%%%%%%%%%%%%%%%%%%%%%%%%%%%%%%%%%%%%%%%%%%%%%%%%%%%%%%%%%%%%%%%%%%%%%%FIG.3
\begin{figure}[!htp]
  \vskip -0.1cm \centerline{
   \hskip -0.3cm
\epsfig{file=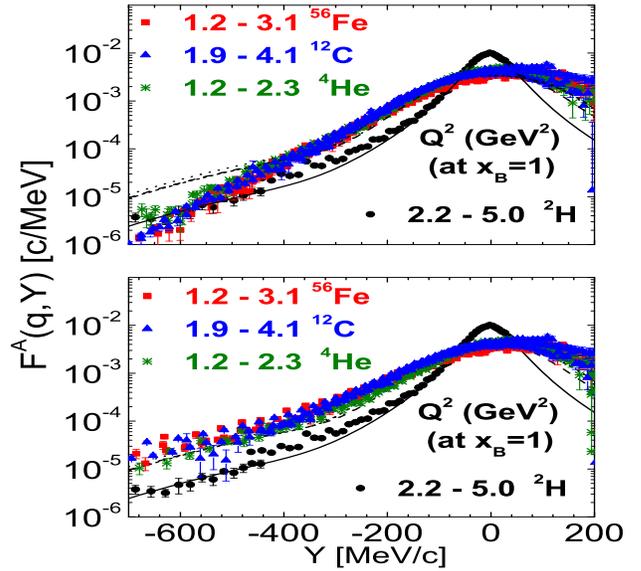,width=12cm,height=8cm}}
  \vskip -0.3cm \caption{The experimental scaling function
  %$F^A_{ex}(q,Y)$
  %(Eq. (\ref{effeexp}),
  (symbols)
for $^4He$,
  $^{12}C$, and $^{56}Fe$  obtained from the experimental data of
  Refs. \cite{deuteron,arrington}. The upper panel shows $F^A(q,Y=y)$
  and the lower panel
  $F^A(q,Y=y_{CW})$. The full, long-dashed, dashed  and dotted curves represent
   the longitudinal momentum
  distributions $f^A(Y)=2\pi \int\displaylimits_{|Y|}^{\infty}n^A(k)kdk$
  for $^{2}H$,
  $^4He$,  $^{12}C$ and  $^{56}Fe$, respectively, calculated with realistic wave functions.}
  \label{Fig3}
\end{figure}
%%%%%%%%%%%%%%%%%%%%%%%%%%%%%%%%%%%%%%%%%%%%%%%%%%%%%%%%%%%%%%%%%%%%
%%%%%%%%%%%%%%%%%%%%%%%%%%%%%%%%%%%%%%%%%%%%%%%%%%%%%%%%%%%%%%%%%%%%%%%%%%%%%%%%%%%%%%FIG.4
\begin{figure}[!htp]
%\vskip 1.0cm
 \centerline{
% \epsfysize=0.44
 %\textwidth\epsfbox{Fig4Chiara.eps}}
 \hskip -0.5cm
    \epsfig{file=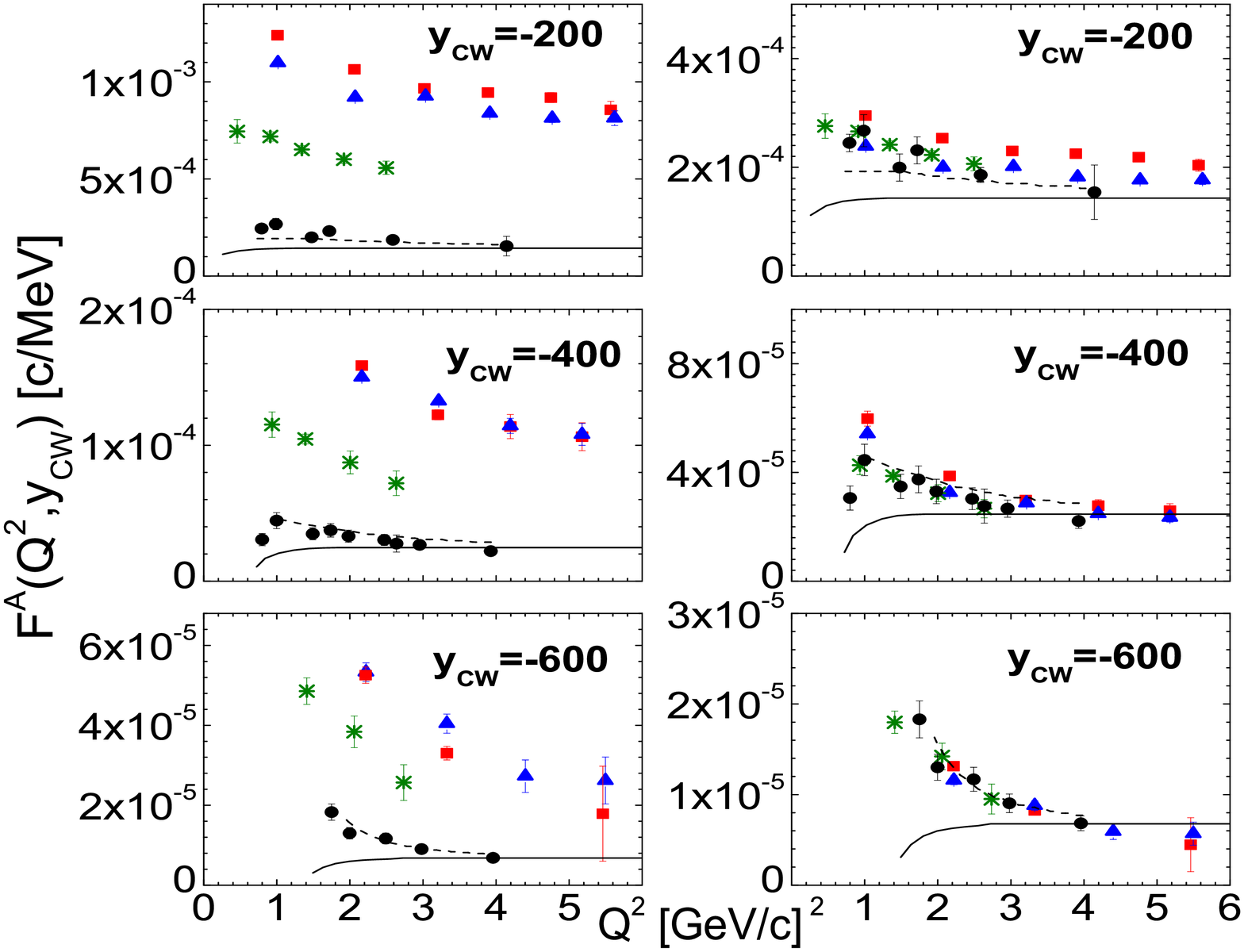,width=8.5cm,height=9.5cm}}
  \vskip -0.3cm \caption{The scaling function $F^A(Q^2,y_{CW})$ from   the lower panel
  of Fig. \ref{Fig3} plotted  {\it vs} $Q^2$
    at fixed values of $y_{CW}$ ($^4He$-{\it asterisks }, $^{12}C$-{\it triangles},
  $^{56}Fe$-{\it squares}).
    In the right panel the data for $^4He$, $^{12}C$ and $^{56}Fe$
  have been divided by the constants $C_4=2.7$,
  $C_{12}=4.0$ and $C_{56}=4.6$, respectively. The theoretical curves refer to $^2H$ and
  represent
  the PWIA results ({\it full}) and the results that include the FSI
  ({\it dashed}), both obtained with the AV18 interaction \cite{av18}.
  Scaling variables are in MeV/c.}
  \label{Fig4}
\vskip -0.5cm
\end{figure}
%%%%%%%%%%%%%%%%%%%%%%%%%%%%%%%%%%%%%%%%%%%%%%%%%%%%%%%%%%%%%%%%%%%%%%%%%%%
%%%%%%%%%%%%%%%%%%%%%%%%%%%%%%%%%%%%%%%%%%%%%%%%%%%%%%%%%%%%%%%%%%%%%%%%%%%
Let us now illustrate the merits of $y_{CW}$-scaling and its
practical usefulness. The main merit is that, because of the
definition of $y_{CW}$,  binding effects play a minor role, as
clearly illustrated
 in Fig. \ref{Fig2};
 practically  $k_{min}(q,\nu,E) \simeq
\vert y_{CW} \vert$ and $B^A(q,y_{CW})\simeq 0$, with  two
relevant
 consequences:
i) to a large extent  $F^A(q,y_{CW}) \simeq f^A(y_{CW})$ [cf. Eq.
(\ref{fqY})], and  ii) as a result of i),  one would expect that
at high values of $y_{CW}$, $F^A(q,y_{CW})$  will behave in the
same way in the deuteron  and in complex nuclei,  since $n^A(k)
\simeq C_{A} n^D(k)$ and, accordingly,  $ f^A(y_{CW}) \simeq C_A
f^D(y_{CW})$; at low values of $y_{CW}$, in contrast,
$F^A(q,y_{CW})$ should exhibit an A dependence generated by the
different asymptotic behavior of the nuclear wave functions in
configuration space. This is fully confirmed in Fig. \ref{Fig3}
which, moreover,  also shows that whereas $F^A(q,y)$ scales
 to a quantity that strongly differs from the longitudinal momentum distribution,
 $F^A(q,y_{CW})$ scales exactly to $f^A(y_{CW})$. This is even better demonstrated in
 Fig. \ref{Fig4}, where
 the effects of FSI are also illustrated. The left
panel shows that (i) scaling is violated  and approached from the
top (which is  clear signature of the breaking down  of the PWIA,
which has to approach scaling from the bottom \cite{ciofi}) and (ii)
the $Q^2$ dependence of the scaling violation appears to be the
same for the deuteron and complex nuclei, a fact that has never
been demonstrated before and represents, in our opinion, a
relevant finding.  To better validate point ii), we have divided
$F^A(Q^2,y_{CW})$ by a constant $C_A$, such as to obtain
$F^A(Q^2,y_{CW})/C_A \simeq F^D(Q^2,y_{CW})$. The results are
shown in the right panel of Fig. \ref{Fig4};  it can again be seen
that not only at high values of $|y_{CW}|$ do all scaling functions
scale in A, but, more importantly, the constants $C_A$
 agree, within the statistical errors, with the
theoretical predictions of Ref. \cite{mark1}, as well as  with the
experimental results on the ratio $R(x_{Bj},Q^2) = 2\sigma_2^A
(x_{Bj},Q^2)/A \sigma_2^D (x_{Bj},Q^2)$ \cite{ratioAD}. The main
findings of our analysis can be summarized as follows: (i) at high
values of $|y_{CW}|$ ( $\gtrsim 200-300 \,MeV/c$) the scaling
function $F^A(Q^2,y_{CW})$  scales to the one of the deuteron,
with scaling constants $C_A$ in qualitative agreement with
theoretical predictions and other types of experimental analysis;
this kind of A-scaling is entirely due to the scaling of the
momentum distributions, $n^A(k) \simeq C_A n^D(k)$, at $k\gtrsim
1.5-2\,fm^{-1}$, which can therefore be investigated by
$y_{CW}$-scaling analysis of inclusive data, owing to the direct
link between the  scaling function $F^A(Q^2,y_{CW})$ and the
longitudinal momentum distributions; ii) the FSI has relevant
effects on the scaling functions up to $Q^2 \simeq 4-5\,GeV^2$
but, most importantly and surprisingly, it exhibits a similar
$Q^2$ dependence in complex nuclei and in the deuteron; this has
neither been observed nor theoretically predicted previously; in a
forthcoming paper it will indeed be shown that the effects of the
FSI on the momentum distribution of a  correlated nucleon are
similar in the deuteron and in a complex nucleus (for preliminary
results see Ref. \cite{ICTP08}).

Useful discussions with M. Alvioli, D. Day, L. Kaptari,  S.
Scopetta, and M. Strikman   are gratefully acknowledged.

\end{document}